# Photonic single sideband RF generator based on an integrated optical micro-ring resonator

Xingyuan Xu, *Student Member, IEEE,* Jiayang Wu, *Member, IEEE,* Mengxi Tan, Thach G. Nguyen, Sai T. Chu, Brent E. Little, Roberto Morandotti, *Fellow, OSA, Senior Member, IEEE,* Arnan Mitchell, *Member, IEEE, and* David J. Moss, *Fellow, OSA, Fellow, IEEE*

*Abstract*—We demonstrate narrowband orthogonally polarized optical RF single sideband generation as well as dual-channel equalization based on an integrated dual-polarization-mode high-Q microring resonator. The device operates in the optical communications band and enables narrowband RF operation at either 16.6 GHz or 32.2 GHz, determined by the free spectral range and TE/TM mode interval in the resonator. We achieve a very large dynamic tuning range of over 55 dB for both the optical carrier-to-sideband ratio and the dual-channel RF equalization.

## I. Introduction

Photonic radio frequency (RF) signal processing [1-4] has attracted great interest for a wide range of applications in radar and communications due to its high performance, including broad bandwidth, low loss, high versatility and reconfigurability, and strong immunity to electromagnetic interference. To date, many key functions have been realized, including advanced modulation format generators [5-10], beamforming [11], transversal filters and signal processors [12-14], RF channelizers [15], optical RF spectrum analyzers [16-18], and many others.

In RF photonic systems, the modulation formats of optical RF signals directly impact their transmission capacity and spectral efficiency and so are key to the design of modern optical RF transmitters [19]. Optical single sideband (OSSB) modulation is an advanced modulation format that has attracted great interest over the past two decades due to its enhanced spectral efficiency and ability to overcome dispersion-induced distortion. Recently, orthogonally polarized OSSB modulation, where the optical carrier and the sideband are orthogonally polarized, has been proposed to further enhance the spectral efficiency and flexibility. Orthogonally polarized OSSB modulation also enables separate manipulation of the optical carrier and sideband via polarization-sensitive optical components, and so has been widely employed in applications ranging from antenna beamforming to microwave photonic signal processing [19-23]. To achieve orthogonally polarized OSSB modulation, many approaches have been demonstrated, including those based on an acousto-optic modulator [23], a differential group delay module [22], a Sagnac-loop-based modulator [24], a dual-polarization quadrature phase shift keying modulator [25], and polarization modulators [26,27]. Fiber-based stimulated Brillouin scattering has also been employed to control the polarization of optical signals for orthogonally polarized OSSB generation [28,29]. Many of these approaches, however, face limitations of one form or another. RF hybrid couplers introduce a bandwidth bottleneck for the whole system, while bulky components and fiber-based structures face challenges in terms of size and stability for practical applications beyond the laboratory.

RF equalizers, which can compensate for imbalance in terms of the frequency response of passive components as well as the gain profile of RF amplifiers [30,31], are also key components in RF systems. Considerable effort has been devoted towards achieving RF equalizers based on state-of-the-art substrate-integrated waveguides [32,33]. However, these approaches can suffer from limitations such as the intrinsic electronic bandwidth bottleneck, limited resolution for equalization (i.e., the minimum equalizing RF bandwidth), lack of tunability of the RF frequency, and limited dynamic range in the equalizer's transmission spectra. Thus, for advanced RF systems requiring a large-dynamic-range, high-resolution, and wideband RF equalization, further advances are required.

Integrated microwave photonics is an attractive approach to address the above challenges [34], with the added benefits of enabling devices with compact footprint, low power consumption, high reliability, and that are mass producible [35]. In this paper, we demonstrate an orthogonally polarized OSSB generator and dual-channel RF equalizer based on an integrated dual-polarization micro-ring resonator (DP-MRR). A spectral interval of ~16.6 GHz and ~32.2 GHz between the TE and TM polarized resonances of the DP-MRR is achieved. Lithographic control of the effective refractive indices of the orthogonally polarized resonant modes of the DP-MRR enable the design and fabrication of waveguides with nearly symmetric waveguide cross section that support both TE and TM polarization modes, while still achieving relatively wide spectral intervals

This work was supported by the Australian Research Council Discovery Projects Program (No. DP150104327). RM acknowledges support by the Natural Sciences and Engineering Research Council of Canada (NSERC) through the Strategic, Discovery and Acceleration Grants Schemes, by the MESI PSR-SIIRI Initiative in Quebec, and by the Canada Research Chair Program. He also acknowledges additional support by the Government of the Russian Federation through the ITMO Fellowship and Professorship Program (grant 074-U 01) and by the 1000 Talents Sichuan Program in China. Brent E. Little was supported by the Strategic Priority Research Program of the Chinese Academy of Sciences, Grant No. XDB24030000.

X. Y. Xu, J. Y. Wu, M. X. Tan, and D. J. Moss are with Centre for Micro-Photonics, Swinburne University of Technology, Hawthorn, VIC 3122, Australia. X. Y. Xu and J. Y. Wu contribute equally to this paper. (Corresponding e-mail: dmoss@swin.edu.au).

T. G. Thach and A. Mitchell are with the School of Engineering, RMIT University, Melbourne, VIC 3001, Australia.

S. T. Chu is with Department of Physics and Material Science, City University of Hong Kong, Tat Chee Avenue, Hong Kong, China.

B. E. Little is with State Key Laboratory of Transient Optics and Photonics, Xi'an Institute of Optics and Precision Mechanics, Chinese Academy of Science, Xi'an, China.

R. Morandotti is with INSR-Énergie, Matériaux et Télécommunications, 1650 Boulevard Lionel-Boulet, Varennes, Québec, J3X 1S2, Canada, with National Research University of Information Technologies, Mechanics and Optics, St. Petersburg, Russia, and also with Institute of Fundamental and Frontier Sciences, University of Electronic Science and Technology of China, Chengdu 610054, China.

of ~16.6 GHz or ~32.2 GHz between the TE and TM polarized resonances in the optical communications band. Hence, at the drop-port, the optical carrier and sideband can be dropped by orthogonally polarized resonances to achieve orthogonal polarized OSSB modulation. Further, at the through-port the notches enable dual-channel RF filtering via phase-to-intensity modulation conversion. Finally, by controlling the polarization angle, we achieve a large dynamic tuning range of over 55 dB for both the optical carrier-to-sideband ratio of the OSSB signal, as well as the extinction ratio of the dual-channel RF equalization.

## II. INTEGRATED DUAL-POLARIZATION-MODE MICRORING RESONATOR

Figure 1(a) shows a schematic of the DP-MRR used in our experiment, which was fabricated on a high-index doped silica glass platform using CMOS compatible fabrication processes [36-41]. High-index (n = ~1.70 at 1550 nm) doped silica glass films were deposited using standard plasma enhanced chemical vapour deposition, then patterned photo-lithographically and etched via reactive ion etching to form waveguides with exceptionally low surface roughness. Finally, silica glass (n = ~1.44 at 1550 nm) was deposited as an upper cladding. The radius of the DP-MRR was ~592 μm, corresponding to a free spectral range (FSR) of ~0.4 nm (~49 GHz). The through-port insertion loss was ~1.5 dB after being packaged with fibre pigtails via butt coupling. The waveguide cross-section was nearly symmetric, with a dimension of 1.5 μm × 2 μm, designed to enable the MRR to support both TE and TM modes. The calculated TE and TM mode profiles are shown in Figs. 1(b) and (c), the corresponding effective indices of which were $n_{eff\_TE}$ = 1.627 and $n_{eff\_TM}$ = 1.624, resulting in slightly different FSRs for the TE and TM resonances and a wide TE/TM separation of $\delta_{TE//TM}$=~16.6 GHz (Fig. 2) near 1550nm. The orthogonally polarized resonances show a similar full width at half maximum (FWHM) of ~ 140 MHz, corresponding to a Q factor of over 1.2 million [42-46], and a 20 dB bandwidth of ~ 1 GHz (Fig. 2(b)). The high Q factor of the MRR enables narrowband RF operation and provides a steep slope for optical filtering, thus enabling RF operation down to 500MHz with 20 dB suppression of the undesired sidebands. Different RF operating frequencies can be realized by designing the TE/TM mode interval via lithographic control of the waveguide cross-section and ring radius, and so the 16.6-GHz spectral interval does not pose a limitation on the RF operation frequency of the orthogonally polarized OSSB generator or the RF equalizer. Further, by employing other adjacent resonances, an operation frequency of 32.4 GHz (= FSR − $\delta_{TE//TM}$) can also be achieved. More generally, MRRs with FSRs of 200 GHz and higher [47] can be realized, enabling even higher RF operation frequencies.

For an arbitrary fixed optical carrier wavelength, the DP-MRR can be thermally tuned to match the carrier wavelength. Thermal tuning can readily be achieved within a resolution of 0.01 °C or even lower, corresponding to MHz level resolution of the DP-MRR, with a millisecond thermal response time [48]. We measured the transmission spectra of the DP-MRR with varying chip temperature to demonstrate this thermal tuning. Figures. 3(a) and (b) show the transmission spectra when the chip temperature changed from 23 °C to 30 °C, and the corresponding two TM resonances (spanning one FSR) and one TE resonance (marked as "TM1", "TE" and "TM2" in Fig. 3(b), respectively). As shown in Fig.

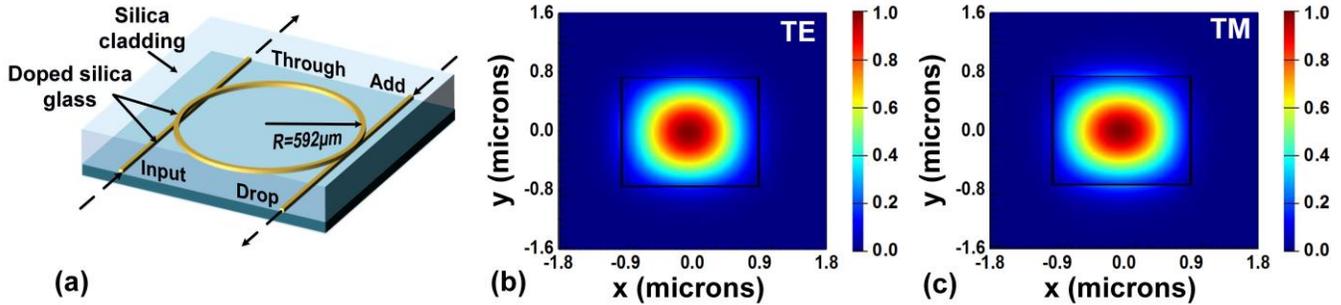

Fig. 1. (a) Schematic illustration of the DP-MRR. (b) TE and (c) TM mode profiles of the DP-MRR.

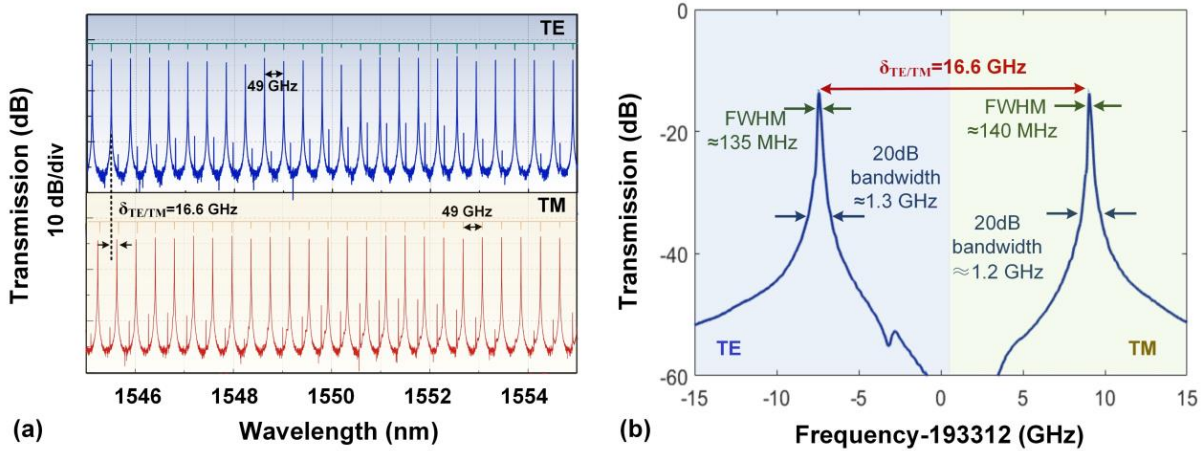

Fig. 2. Measured transmission spectra of the on-chip 49GHz-FSR MRR. (a) Through-port (cyan, yellow) and drop-port (blue, red) transmission spectra of TE and TM polarization. (b) Zoom-in view of the drop-port transmission showing the resonances with full-width at half-maximum (FWHM) of 140 MHz, corresponding to a Q factor of over $1.2\times10^6$.

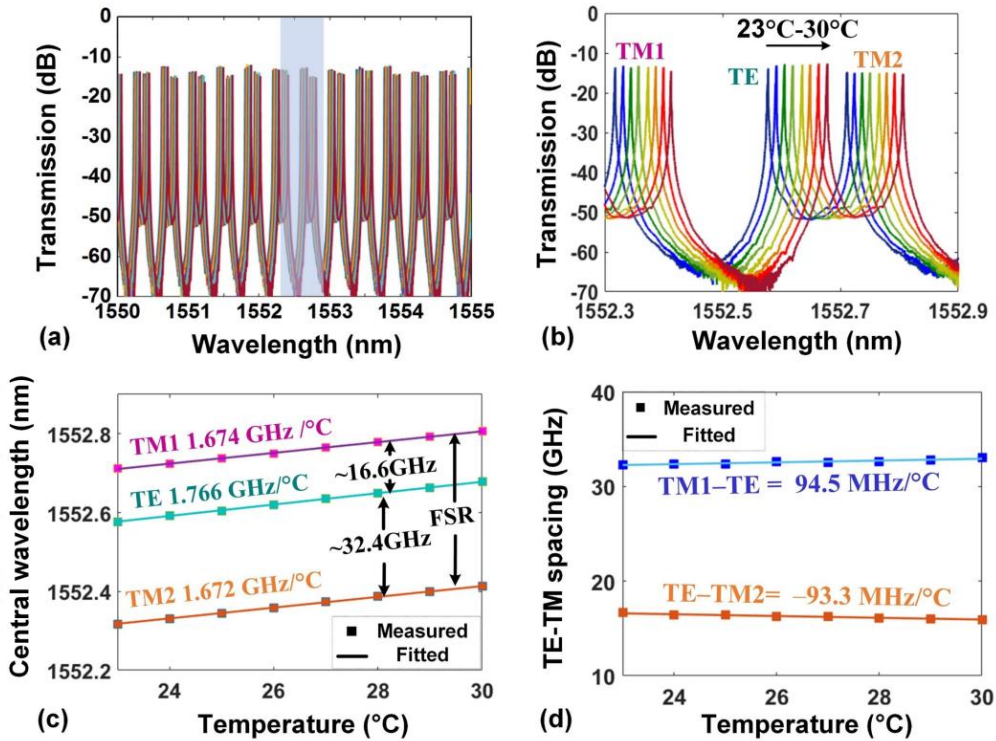

Fig. 3. (a)−(b) Transmission spectra of the OSSB generator with the chip temperatures ranging from 23 °C to 30 °C. Extracted relationship between the chip temperature and (c) central wavelength of the resonances, (d) the spacing between TE and TM resonances.

3(c), the center wavelength of the resonances redshifted at a rate of ~1.67 GHz/°C for the TM resonances and ~1.77 GHz/°C for the TE resonances. Thus the TE/TM mode interval $\delta_{TE//TM}$ varies slightly with temperature. Figure 3(d) shows the TE/TM mode interval between the resonances marked as "TE", "TM1" and "TM2", as the chip temperature was varied. The ~90 MHz/°C slope is much smaller than the 20 dB bandwidth of the MRR (~1 GHz) and in any event can be compensated for via waveguide design (dispersion engineering).

### III. ORTHOGONALLY POLARIZED OPTICAL SINGLE SIDEBAND GENERATOR

Figure 4 shows a schematic of the orthogonally polarized OSSB generator. As mentioned, the DP-MRR was fabricated to have a nearly square cross-section to support both TE and TM polarized modes while still yielding a significant difference in effective indices between the TE and TM modes, leading to a resonance shift and slightly different FSRs between the two polarizations. Next, a continuous-wave (CW) light from a tunable laser source was modulated to generate a double sideband (DSB) signal, which was then fed into the DP-MRR with the polarization aligned to 45° relative to the TE-axis (Fig. 4(i)). When the wavelength of the CW light and RF frequency matched the orthogonally polarized resonances of the DP-MRR, the optical carrier and one of the sidebands of the generated DSB signal could be dropped by the TE and TM resonances, respectively, thus achieving orthogonally polarized OSSB modulation (Fig. 4(ii)). Moreover, when the orthogonally polarized OSSB signal went through an optical polarizer, the ratio of TE to TM light could be adjusted via polarization control, and so single-polarization OSSB modulation with a tunable optical carrier to sideband ratio (OCSR) was achieved.

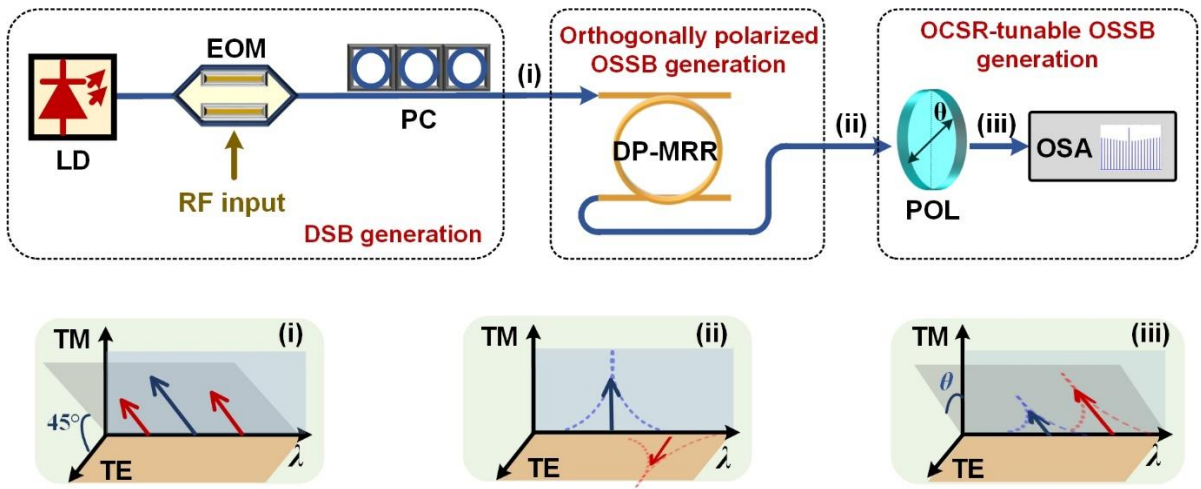

Fig. 4. Schematic of the proposed orthogonally polarized optical single sideband (OSSB) generator. LD: laser diode. EOM: electro-optical modulator. PC: polarization controller. POL: optical polarizer. OSA: optical spectrum analyzer. DP-MRR: dual-polarization-mode micro-ring resonator. DSB: double sideband. OCSR: optical carrier to sideband ratio. (i) The 45° polarized carrier is modulated with dual side-bands. (ii) The carrier is passed by the TM resonance of the DP-MRR and the upper sideband is passed by the TE resonance of the DP-MRR while the lower sideband is rejected by the DP-MRR. (iii) A polarizer extracts the 45° components of both carrier and upper sideband, projecting the SSB signal onto a single polarization.

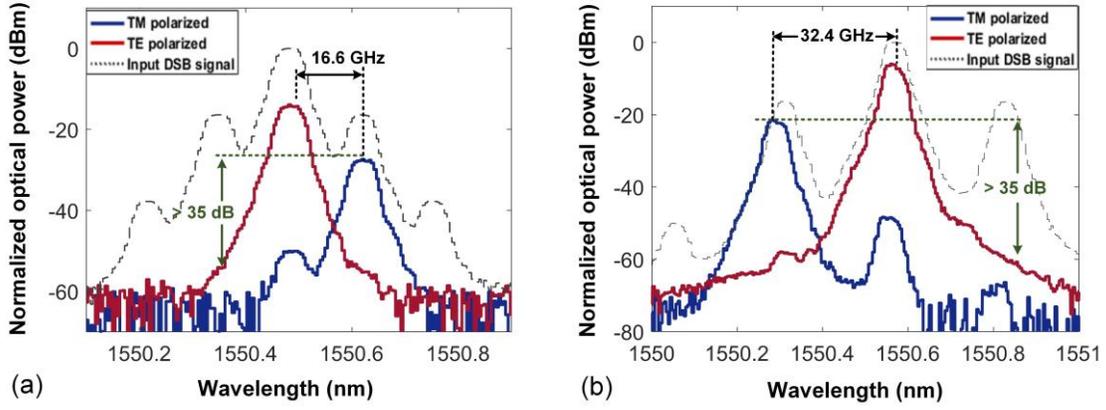

Fig. 5. Optical spectra of the generated orthogonally polarized OSSB signal.

The difference in effective indices of the orthogonal polarizations resulted in a strong polarization dependence in the transmission of the DP-MRR. To analyze our device we used the Jones matrix formalism, where the polarization eigenmodes of the DP-MRR serve as a natural basis, and the drop-port transmission of the DP-MRR can be written as

$$R = \begin{pmatrix} D_{TE} & 0 \\ 0 & D_{TM} \end{pmatrix} \quad (3.1)$$

where $D_{TE}$ and $D_{TM}$ are the drop-port transfer functions of TE and TM modes given by

$$D_{TE} = \frac{-k^2 \sqrt{a} e^{i\phi_{TE}/2}}{1 - t^2 a e^{i\phi_{TE}}} \quad (3.2)$$

$$D_{TM} = \frac{-k^2 \sqrt{a} e^{i\phi_{TM}/2}}{1 - t^2 a e^{i\phi_{TM}}} \quad (3.3)$$

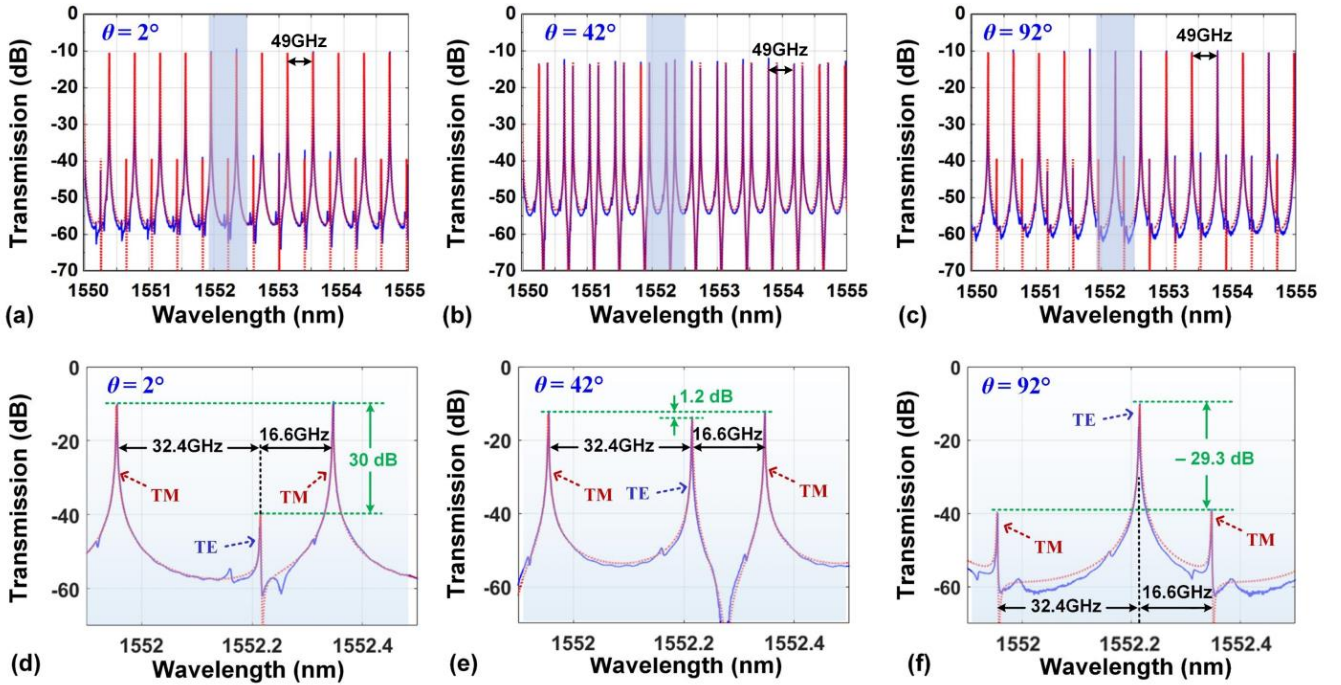

Fig. 6. Transmission spectra of the OSSB generator with (a) $\theta = 2°$, (b) $\theta = 42°$, (c) $\theta = 92°$, where $\theta$ denotes the polarization angle as shown in Figure 4. (d)–(f) show zoom-in views of the shaded areas in (a)–(c), respectively.

where *t* and *k* are the transmission and cross-coupling coefficients between the bus waveguide and the micro-ring ($t^2 + k^2 = 1$ for lossless coupling), *a* represents the round-trip transmission factor, $\phi_{TE} = 2\pi L \times n_{eff\_TE} / \lambda$ and $\phi_{TM} = 2\pi L \times n_{eff\_TM} / \lambda$ are the single-pass phase shifts of TE and TM modes, respectively, with *L* denoting the round-trip length, $n_{eff\_TE}$ and $n_{eff\_TM}$ denoting the effective indices of TE and TM modes, and $\lambda$ denoting the wavelength.

The Jones matrix of the polarizer is

$$P(\theta) = \begin{pmatrix} \sin^2\theta & \cos\theta\sin\theta \\ \sin\theta\cos\theta & \cos^2\theta \end{pmatrix} \quad (3.4)$$

where $\theta$ is the polarization angle between the polarizer's direction and the TM axis. For a general input $E_0 \begin{bmatrix} \cos 45° \\ \sin 45° \end{bmatrix}$, the output field after the polarizer is $P(\theta) R E_0 \begin{bmatrix} \cos 45° \\ \sin 45° \end{bmatrix}$, corresponding to an output intensity given by [51]

$$I(\theta) = \frac{E_0^2}{2}\left[|D_{TE}|^2 \cdot \sin^2\theta + |D_{TM}|^2 \cdot \cos^2\theta + |D_{TE}| \cdot |D_{TM}| \cdot \sin 2\theta \cdot \cos(\varphi_{TE} - \varphi_{TM})\right] \quad (3.5)$$

where the input is $\varphi_{TE}$ and $\varphi_{TM}$ are the complex phase angle of $D_{TE}$ and $D_{TM}$.

According to the above equation, the output optical power dropped from the TM and TE resonance is proportional to $\cos^2\theta$ and $\sin^2\theta$, i.e., the loss induced by the polarization conversion is proportional to $\sin^2\theta$ and $\cos^2\theta$, respectively. Particularly, when $\theta = 45°$, the loss induced by the polarization conversion is 3 dB for both TE and TM modes. The OCSR (with the TM resonance for the carrier and the TE resonance for the upper sideband) is given by

$$\text{OCSR}(\theta) \propto \cot^2\theta \quad (3.6)$$

which can be continuously tuned via changing $\theta$. Moreover, since $\cot^2\theta$ can infinitely approach 0 or 1 as $\theta$ approaches $\pi/2$ or 0, an ultra-large

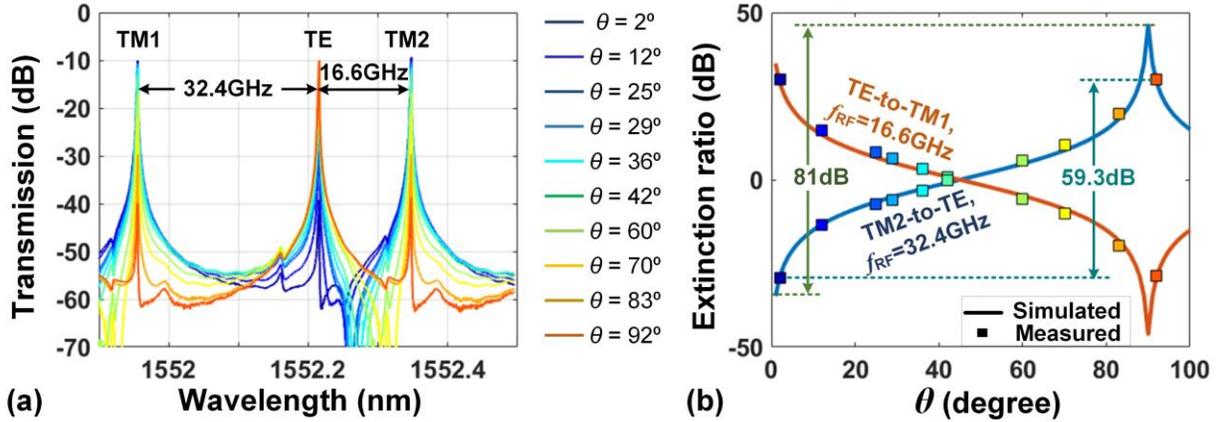

Fig. 7. (a) Transmission spectra of the OSSB generator with $\theta$ varying from 2° to 92°, and (b) extracted extinction ratio between the TE and TM resonances labeled as "TE", "TM1" and "TM2" in Fig. 7(a), corresponding to 16.6 GHz and 32.4 GHz RF operation, respectively.

dynamic tuning range of the OCSR can be achieved.

During the experiment, we tuned the optical wavelength into the TE resonance centered at 1550.47 nm, and drove an intensity modulator (ixBlue) with 16.6 GHz / 32.4 GHz RF signals, such that the upper/lower sideband could be dropped by the adjacent TM resonance on the carrier's red/blue side. An orthogonally polarized carrier and sideband were obtained at the drop-port of the DP-MRR, where the optical power of the unused sideband was suppressed by over 35 dB as compared with the used sideband, as shown in Fig. 5.

Next, the orthogonally polarized OSSB signal was converted to a single-polarization OSSB signal via a polarizer, and a tunable OCSR was achieved by adjusting the polarizer angle. We measured the transmission spectra of the established OCSR-tunable OSSB generator, as shown in Fig. 6. As $\theta$ was changed from 2° to 92°, the extinction ratio between TM and TE resonances varied from 30 dB to –29 dB, corresponding to an OCSR tuning range of up to 59.3 dB. Figures. 6(d) – (f) show details of the varying extinction ratio and clearly indicate that up to 32.4 GHz RF operation is within the capability of our OCSR-tunable OSSB generator. Figure 7 plots transmission spectra and corresponding extracted extinction ratio for both 16.6 GHz and 32.4GHz RF operation as the polarization angle changed from 2° to 92°, which matches very well with theory. We note that an ultra-large dynamic range of > 80 dB is predicted in Fig. 7(b), which can be achieved by adjusting $\theta$ with finer resolution.

Figure 8 shows the optical spectra of the generated 16.6 GHz and 32.4 GHz single-polarization OSSB signals with a continuously tunable OCSR ranging from −22.7 to 41.4 dB and −27.1 to 52.2 dB, respectively, verifying the feasibility and high performance of our OCSR-tunable OSSB generator. Finally, by employing carrier to sideband shifts of multiple FSRs, yet higher RF frequencies can be realized from the same device. For the device considered here, these RF frequencies correspond to 65.6 GHz = 16.6 GHz + FSR and 81.4 GHz = 32.4 GHz + FSR and so on.

We note that the high-Q MRR provided a high RF selectivity for the OSSB generator, thus enabling self-oscillating high-frequency RF sources by means of optoelectronic oscillators which typically feature ultra-low phase noise. This is an important approach for long-distance RF standards delivery due to the fact that the OSSB modulation format is immune to dispersion-induced RF power fading, thus enabling a wide range of applications from radio astronomy telescope arrays (such as the Atacama Large Millimeter Array) to the test of fundamental physics.

## IV. RF EQUALIZER

Figure 9 shows a diagram of the photonic RF equalizer based on the DP-MRR. A CW light from a tunable laser source was phase modulated by an input RF signal to generate counter-phase sidebands, with the polarization aligned to have an angle of $\theta$ to the TE-axis (Fig. 9(i)). Then the TE and TM components of the phase-modulated signal were filtered by the orthogonally polarized resonances (notches) of the DP-MRR, separately (Fig. 9(ii)), where the imbalance between the counter-phase sidebands was introduced to achieve phase-to-intensity modulation conversion. After that, the filtered optical signals featuring orthogonal polarization states were converted into RF signals and combined upon photodetection. Thus, the high-Q orthogonally polarized optical resonances were mapped onto the RF domain (Fig. 9(iii)), achieving a high frequency selective RF filter with dual passbands. The bandwidth of the RF filter is determined by the Q factor of the DP-MRR, while the centre frequencies are determined by the relative spacing between the optical carrier and adjacent resonances. By controlling the polarization angle $\theta$, the proportion of TE- and TM-polarized light could be adjusted continuously, and thus after optical-to-RF mapping [50] the extinction ratio between the dual RF passbands could be tuned to achieve RF equalization.

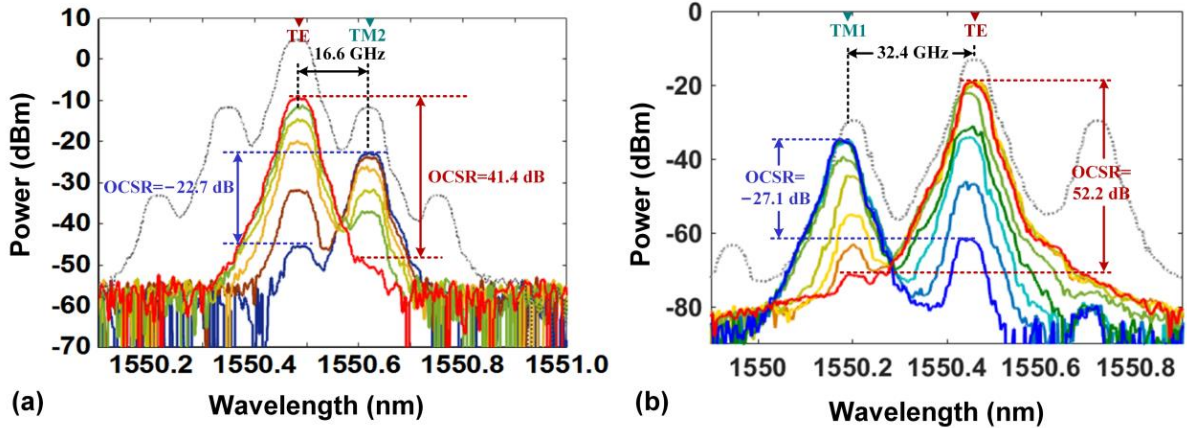

Fig. 8. Optical spectra of the generated OSSB signal with continuously tunable optical carrier-to-sideband ratio (OCSR) driven by (a) 16.6 GHz and (b) 32.4 GHz RF signal. The optical carrier was dropped by the "TE" resonance, the 16.6 GHz and 32.4 GHz RF sideband were dropped by the "TM2" and "TM1" resonances, as marked in the figure.

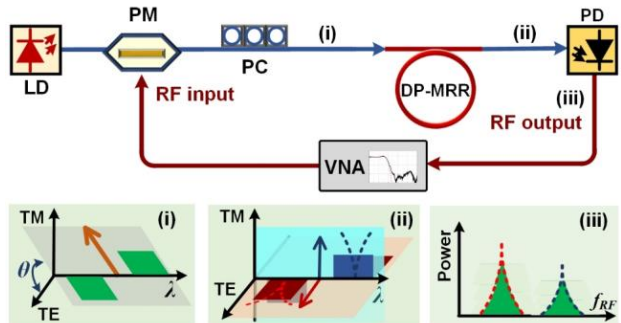

Fig. 9. Schematic diagram of the photonic RF equalizer based on a dual-polarization-mode micro-ring resonator. LD: laser diode. PM: phase modulator. PC: polarization controller. DP-MRR: dual-polarization-mode micro-ring resonator. PD: photo-detector. VNA: vector network analyzer.

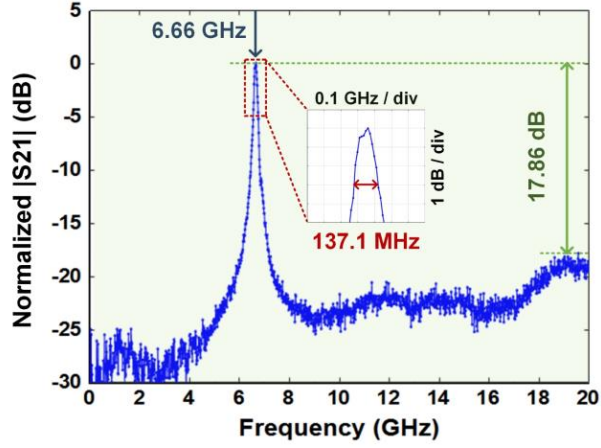

Fig. 10. RF transmission of a single passband with TM-polarized optical input.

The through-port transmission of the DP-MRR can be written as

$$R = \begin{bmatrix} T_{TE} & 0 \\ 0 & T_{TM} \end{bmatrix} \quad (4.1)$$

where $T_{TE}$ and $T_{TM}$ are the through-port transfer functions of the DP-MRR given by

$$T_{TE} = \frac{t(1-ae^{i\phi_{TE}})}{1-t^2 ae^{i\phi_{TE}}} \quad (4.2)$$

$$T_{TM} = \frac{t(1-ae^{i\phi_{TM}})}{1-t^2 ae^{i\phi_{TM}}} \quad (4.3)$$

For a phase modulated optical signal $E_0 \begin{bmatrix} \cos\theta \\ \sin\theta \end{bmatrix}$, the output field of the DP-MRRs is

$$E_{ii} = RE_0 \begin{bmatrix} \cos\theta \\ \sin\theta \end{bmatrix} = E_0 \begin{bmatrix} T_{TE} \cdot \cos\theta \\ T_{TM} \cdot \sin\theta \end{bmatrix} \quad (4.4)$$

As reflected by the above equation, the centre frequencies of RF passbands supported by TE- and TM-resonances are determined by the relative spacing between the optical carrier and the MRR resonances, thus indicating tunable operation regions for the RF equalizer. Moreover, the optical power of the TE- and TM-polarized optical signals is proportional to $\cos^2\theta$ and $\sin^2\theta$, respectively. Thus, after photo-detection the extinction ratio between the RF passbands (corresponding to the DP-MRR's TE- and TM-polarized resonances) is given by

$$ER(\theta) \propto \cot^2\theta \quad (4.5)$$

Similar to the tunable OCSR in OSSB generation, $ER(\theta)$ can also be continuously tuned via changing $\theta$, and since $\cot^2\theta$ can infinitely approach 1 or 0 as $\theta$ approaches 0 or $\pi/2$ (limited only by the polarizer performance), a large tuning range of the extinction ratio can be expected, indicating an ultra-large dynamic tuning range for RF equalization.

During the experiment, we first investigated the resolution and tunability of a single RF passband by setting the input optical signal as TM-polarized ($\theta = 90º$). The RF transmission spectra measured by a vector network analyser (VNA) is shown in Fig. 10. The 3dB-bandwidth of the passband is 137.1 MHz, which defines the resolution of the RF equalizer. The tunability of the passband's centre frequency was achieved by changing the carrier wavelength (Figs. 11(a) – (b)), the chip temperature of the DP-MRR (Figs. 11(c) – (d)), and the input optical power (Figs. 11(e) – (f)). As can be seen, all of the tuning methods can effectively shift the centre frequency of the RF passband with a 3dB-bandwidth of ~140MHz, which can achieve tunability for the high-resolution RF equalizer.

Varying the RF equalizer's operation frequency was obtained by tuning the carrier wavelength (Figs. 12(a) – (b)) and the chip temperature of the MRR (Figs. 12(c) – (d)), continuously covering a range of 14.6 GHz. Extracted centre frequencies of the RF passbands supported by TE- and TM-resonances (i.e., the TE and TM centre frequencies in Figs. 12(b) and (d)) show the efficiency of each tuning method, verifying their effectiveness. Tuning the extinction ratio between the TE- and TM-RF passbands was achieved by adjusting the polarization angle $\theta$ of input light (as indicated in Fig. 9). The optical transmission spectra of the DP-MRR's through-port and drop-port were measured as $\theta$ was varied (as shown in Fig. 13(a)). Due to limited resolution of the tuning angle, the varying TE- and TM-notches of the through-port transmission could not be fully measured, and so we additionally measured the drop-port transmission. As shown in Fig. 13(b), a wide tuning range from −27.4 dB to 28.2 dB, corresponding to a dynamic range of over 55 dB, was achieved for the extinction ratio between the two passbands of the RF dual-channel equalizer, verifying the feasibility and high performance of our approach.

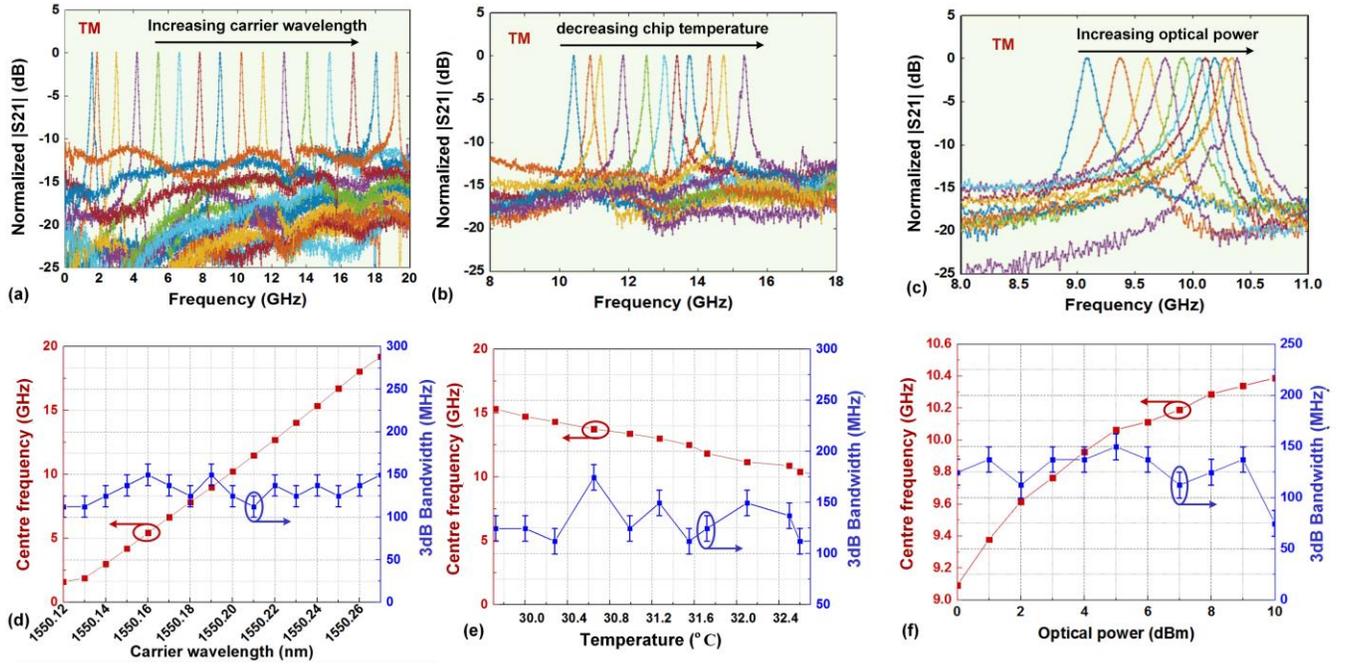

Fig. 11. RF transmission of the single passband with varying (a) carrier wavelength, (b) chip temperature, and (c) input optical power. (d−f) Extracted centre frequency and 3dB bandwidth.

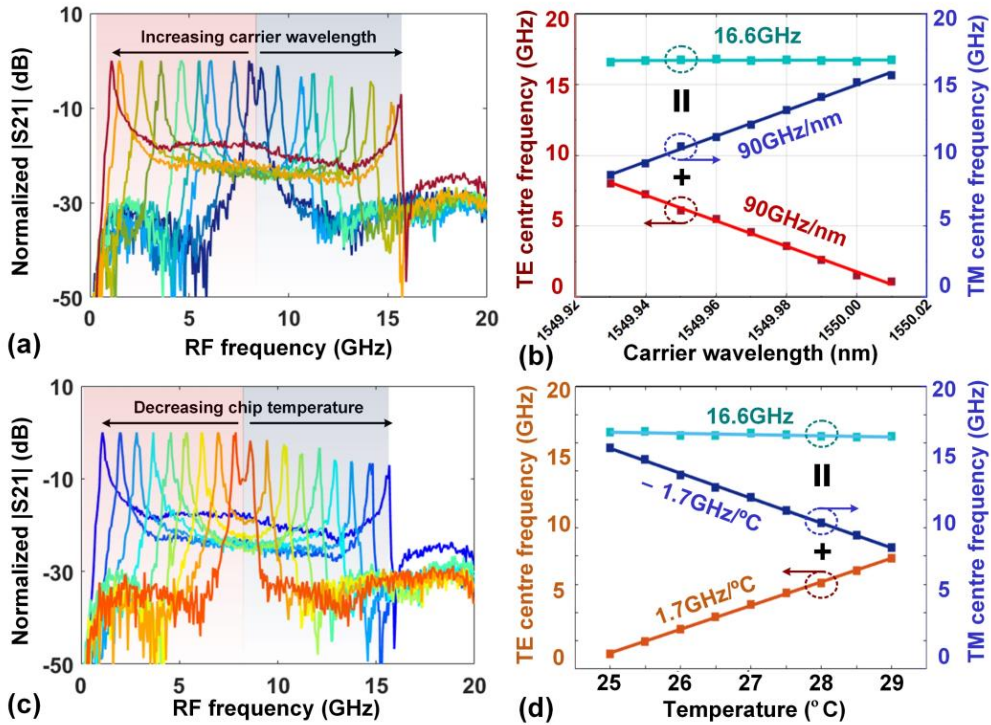

Fig. 12. RF transmission of the proposed equalizer with varying operation frequencies by tuning the (a) carrier wavelength and (c) chip temperature. (c) and (d) Extracted corresponding center frequencies of TE- and TM-passband.

In this work, we focused on achieving narrowband optical single sideband generation and high-resolution RF equalization. For applications

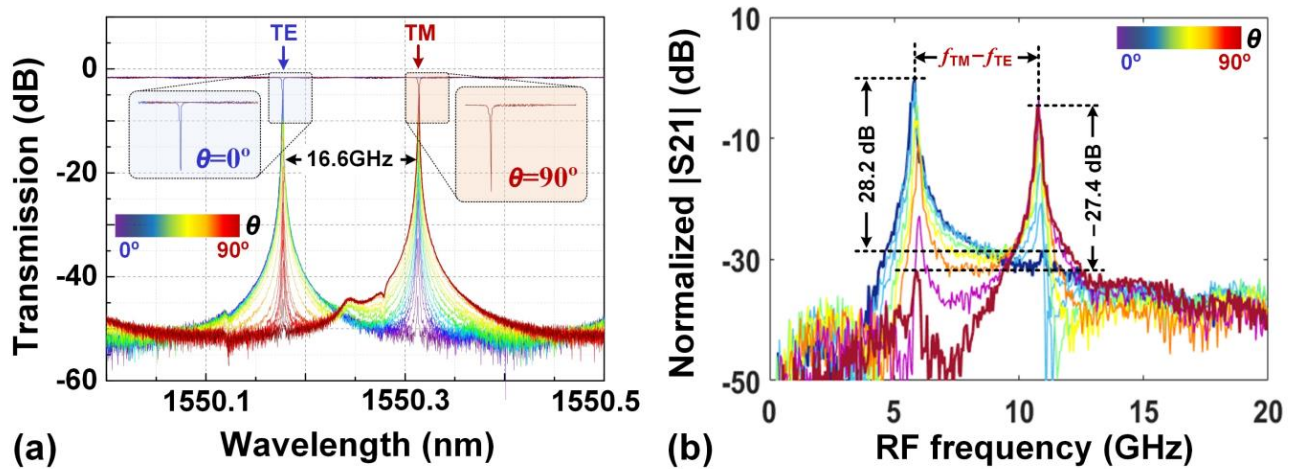

Fig. 13. (a) Optical through-port and drop-port transmission spectra of the DP-MRR and (b) RF transmission of the proposed equalizer with varying extinction ratio between TE- and TM-passband as the polarization angle $\theta$ of input light varies from 0º to 90º. $f_{TM}$–$f_{TE}$ denotes the spacing between the TM-passband and TE-passband, which is wideband tunable (as shown in Fig. 12) and in this plot equals to 4.8 GHz

requiring a broad RF instantaneous bandwidth, either lower Q MRRs [51] or higher order filters [52,53] could be employed to replace the high-Q MRR. The former can yield 3dB bandwidths from 2−12 GHz, corresponding to Q factors ranging from 60,000−10,000, while high order filters can achieve bandwidths of 100GHz or even higher. Finally, for dynamic tunability of the operational RF, an approach based on cascaded MRRs supporting both TE- and TM-polarization modes can be used, where the spectral interval between TE and TM resonances can be tuned via separate thermal control of the MRRs, although this comes at the expense of slightly higher overall loss compared with a single-MRR device.

## V. Conclusion

We demonstrate an orthogonally polarized optical single sideband (OSSB) generator and dual-channel RF equalizer based on an integrated dual polarization micro-ring resonator (DP-MRR). With lithographic control of the effective refractive indices of the orthogonally polarized modes, the TE and TM resonances of the DP-MRR were widely separated with a spacing of 16.6 GHz in the optical communication band, such that at the drop-port, the optical carrier and sideband were separated by the orthogonally polarized resonances to achieve orthogonally-polarized OSSB modulation. At the through-port, the notches enabled dual-channel RF filtering via phase-to-intensity modulation conversion for equalization. By controlling the polarization angle, we achieved a large dynamic tuning range for the optical carrier-to-sideband ratio of the OSSB signal and the dual-channel RF equalization. This approach provides a new way to realize OSSB generation and RF photonic equalization with a reduced footprint and improved performance, which is promising for RF photonic signal processing in radar and communications systems.